# Planar Metasurface Antenna with Tunable via Boundaries for Computational Imaging

Toufiq M. Hossain, *Graduate Student Member, IEEE*, Andrey E. Miroshnichenko, David A. Powell, *Senior Member, IEEE*

*Abstract*— The fusion of metasurface antennas and computational imaging facilitates the design of microwave imaging systems which require no lenses, phase shifters or moving parts. The technique involves the generation of appropriately designed diverse measurement modes to encode the scene information into a small number of measurements. We propose a novel boundary-tunable parallel plate waveguide-based metasurface antenna for computational microwave imaging. The proposed antenna leverages a switchable boundary of two layers of vias, to efficiently change the waveguide modes supported by the antenna cavity, leading to diverse measurement modes in the scene plane. The superiority of the boundary tuning approach over the frequency diversity approach for the same antenna is confirmed using the singular value decomposition. Synthetic imaging is performed using a coupled dipole model, which quantitatively proves the efficacy of the proposed antenna along with the robustness against noise down to 15 dB SNR.

## I. INTRODUCTION

Microwave imaging is a versatile class of imaging methods, with applications ranging from medical diagnostics [1]–[3], synthetic aperture radar [4], remote sensing, nondestructive testing [5], security screening [6], space observation and holographic imaging [7]–[10]. Since microwaves can penetrate optically opaque and non-conducting clothing or materials, they offer non-invasive non-ionizing imaging modalities, and there has been significant interest in developing microwave imaging systems.

Two main architectures are most commonly used for conventional microwave imaging [11]. In synthetic aperture radar (SAR), a single antenna (or pair of transmitting and receiving antenna) performs a mechanical raster scan of the scene while in an electronically scanned array (ESA) multiple transmitting and receiving antennas are used for interrogating the scene. In spite of their ubiquitous applications, ESAs require many RF amplifiers, oscillators and phase shifters and the SAR imaging process requires long data acquisition time due to the mechanical movements of the antennas [12]–[14].

Driven by afore-mentioned technical constraints, researchers have proposed Computational Imaging (CI) strategies for microwave and milli-meter wave frequencies [15]–[21]. This *'conventional to computational imaging'* trend

Corresponding Author: Toufiq Md. Hossain (email: toufiqmhossain@ieee.org.).Toufiq M. Hossain, David A. Powell, and Andrey E. Miroshnichenko are with School of Engineering and Information Technology, University of New South Wales, Canberra. Northcott Drive, Campbell, ACT 2600, Australia

followed similar paradigm shift in optical frequencies after pivotal works of Duarte et al. reporting a *'single-pixel camera'* [22]. The development of compressive sensing [20], [23] provided the algorithms that form the basis for this imaging technique [24]. After the first report of this computational microwave imaging in Ref. [25], several applications have emerged using this technique, including computational phaseless ghost imaging [26], [27], single frequency imaging [28], [29], intelligent imaging [30], and imaging a human at video frame rate [31].

Computational imaging is performed by encoding the scene information into a set of small number of appropriately designed measurement modes, which are typically applied sequentially. In this technique, precisely defined antenna fields and a model of scattering from the scene are used to design the measurement modes, allowing a well-defined forward model. This forward model is then used for reconstruction of the scene. The modelling of the scene is normally performed using the first order Born approximation (i.e. neglecting multiple scattering), whereas the aperture fields are numerically described using approximations such as a discrete dipole model or they are measured by near field scanning [25]. The key requirement of this strategy is to generate measurement modes that can sample as distinct information as possible from the scene. In optics, multiplexing hardware such as spatial light modulators, digital micromirror devices [22], [32], [33], coded apertures [34] and random scattering media [35] are used for creating diverse measurement modes, whereas in the microwave regime they are generated using metamaterial antennas capable of generating the measurement modes using the frequency diverse approach [25], [36] or the dynamic tuning approach [12], [37]–[39].

The first reported approach using frequency diversity is known as *frequency diversity by element resonance*. The antenna consists of resonant metamaterial elements which leak energy from the waveguide and the diversity of measurement modes is determined by the elements' frequency responses. It requires maximizing the Q factor of metamaterial elements to reduce the correlation among the measurement modes. However, this high Q factor leads to high electric currents on the complementary metamaterials, leading to high conductive losses. Therefore, the radiation efficiency of the antenna decreases as the Q factor of the metamaterials increases, leading to an unfavorable trade-off between signal to noise ratio and the diversity of measurement modes. To overcome this limitation, *frequency diversity by feed-diversity* was proposed, based on a deformed cavity-backed antenna, taking advantage of the cavity's high Q-factor and its ability to support multiple frequency dependent modes [14]. In these structures, small irises or holes were etched into the upper surface of a hollow deformed cavity. In this approach the diversity of the measurement modes arises from the strong spatial variation in excited waveguide modes when the frequency is swept, due to their high Q-factor. However, since radiation is the primary mechanism of energy loss in this case, the factor that depletes the design space is the *loaded* Q-factor which generally decreases with the increase of the radiating elements as it has been quantifiably described in [40], [41]. Therefore, a trial and error procedure is required to balance the trade-off between the Q-factor and the number of elements [31], [42].

To tackle the stringent trade-off between the Q-factor, radiation efficiency and the number of radiating elements in the frequency diverse approach, dynamic metasurface apertures [12], [15], [43] were proposed for computational microwave imaging. In these approaches switchable elements (PIN diodes or varactor diodes) are used to modify the pattern of aperture illumination. The advantage of this technique is enlarging the design space with help of tunability to generate diverse measurements for each frequency.

The first kind of dynamic metasurface aperture was reported with a 1D linear configuration [12], and later on a two-dimensional *planar* dynamic metasurface was proposed in [43]. These two works can be regarded as alternative pathways to achieve diversity using the *element's response*. The radiation from the metamaterial elements is turned on/off using PIN diodes. The operating principle of this design is described in detail in [15]. However, since the exciting waves dies out soon due to the radiation from the elements, multiple feed were necessary in the 1D case. Moreover, in both designs bias lines must be connected with each of the metamaterial elements distributed across the aperture, resulting in significant fabrication complexity.

On the other hand, an alternative pathway to achieve *feed diversity* that overcomes some of the limitations imposed by the Q-factor, was proposed in [44], based on a cavity with tunable boundary. A tunable impedance surface was used as the bottom surface of the cavity while the top surface was perforated with simple slot structures for radiation. This design is quite promising since it separates the measurement modes' tunability from the radiation phenomenon and thus mode diversity is not strongly reliant on a high Q-factor. In addition, the impinging waves are reflected with different phase profile facilitated by locally tunable elements, leading to diverse waveforms at each frequency. However, this configuration is bulky in size ($7\lambda \times 7\lambda \times 4\lambda$) and is incompatible with planar fabrication techniques.

In this paper we propose a design combining the advantage of a *planar* dynamic approach with the separation of tunability from radiating elements. The basic principle is similar to that reported in [44] of changing the boundary conditions to achieve feed diversity however, using a parallel-plate waveguide-based configuration that can be fabricated using printed circuit technology. The proposed configuration ensures adequate diversity of the measurement modes for CI. Contrary to design of [12], [43], this configuration does not require the bias lines to run through the radiating area to connect with each of the radiating elements. Instead, the biasing lines can be connected with groups of vias in the inner layer of the side boundaries of the cavity, leaving the radiating elements solely for radiation. A similar concept of changing the via boundaries has also been reported in [45] for fixed-frequency beam steering using substrate integrated waveguide architecture. The principle of operation of the antenna, semi-analytical model and imaging using the Two-step Iterative Shrinkage/Thresholding (TwIST) algorithm [46] is presented as proof of the efficacy of the proposed antenna.

## II. DESIGN OF PROPOSED ANTENNA

The proposed antenna consists of a parallel-plate waveguide cavity operating at 8-12 GHz. The top side of the antenna is etched with 180 radiating cELC metamaterial elements. The metasurface antenna is used as a transmitter in the imaging setup while 4 open-ended waveguide antennas are used as receivers. The imaging setup is shown in Fig. 1(a). To cover the full operating bandwidth, three different metamaterial elements are designed (see Fig. 2), with dimensions given in Table I. We have used 60 metamaterial elements of each size and randomly distributed them across the antenna aperture. The size the antenna is 500 ×500×1.27 mm (16.67λ×16.67λ×0.04λ) at center frequency of 10 GHz). For simplicity, in simulations we consider the parallel-plate waveguide to be a lossless medium with permittivity $\epsilon = 1$ and height of 1.27 mm (0.23λ) which allows only the TEM mode to propagate between adjacent metamaterial elements. Moreover, to ensure that there is no strong-near field coupling between them the radiating elements are not placed too closely, enabling their interaction to be described by dipoles coupling to the fundamental waveguide mode. The side boundaries of the waveguide cavity are formed by two layers of vias. The vias are placed at 4.3 mm (0.14λ) distance from each other, and they are of 0.3 mm radius (~λ/10). The outer layer of vias is fixed, while the inner layer (inset by 8.5 mm (~λ/4) distance) is a tunable layer. On each side of the antenna, the inner layer is divided into 28 tunable subgroups consisting of 3 vias, which are switchable between on and off condition. In the *on* condition, a subgroup of the inner layer is connected with both the top and bottom plates of the waveguide cavity, shifting the boundary inwards by a quarter wavelength. In the *off* condition, a subgroup of the inner layer of vias is disconnected from the top and bottom waveguide plates and are assumed to have negligible scattering of the waveguide mode. Therefore, the waves scattered from each of the outer vias traverses an electrical distance of half wavelength from the inner layer to the outer layer and back. In this way switching between reflection from the inner layer and the outer layer mimics the behavior of switching between PEC and PMC boundary of our previously reported antenna [38]. For other incidence angle the waves are reflected with different phase from the outer fixed boundary, leading to spatially varying speckle inside the cavity which is the primary objective for computational imaging. By pseudo-randomly switching the inner subgroups between on and off conditions, the waveguide modes can be changed inside the parallel plate waveguide which is sampled by the metamaterial elements and as a result the generated radiation pattern also changes. The set of switching states is initially chosen randomly but is recorded and kept consistent throughout the measurement to enable reconstruction of the image.

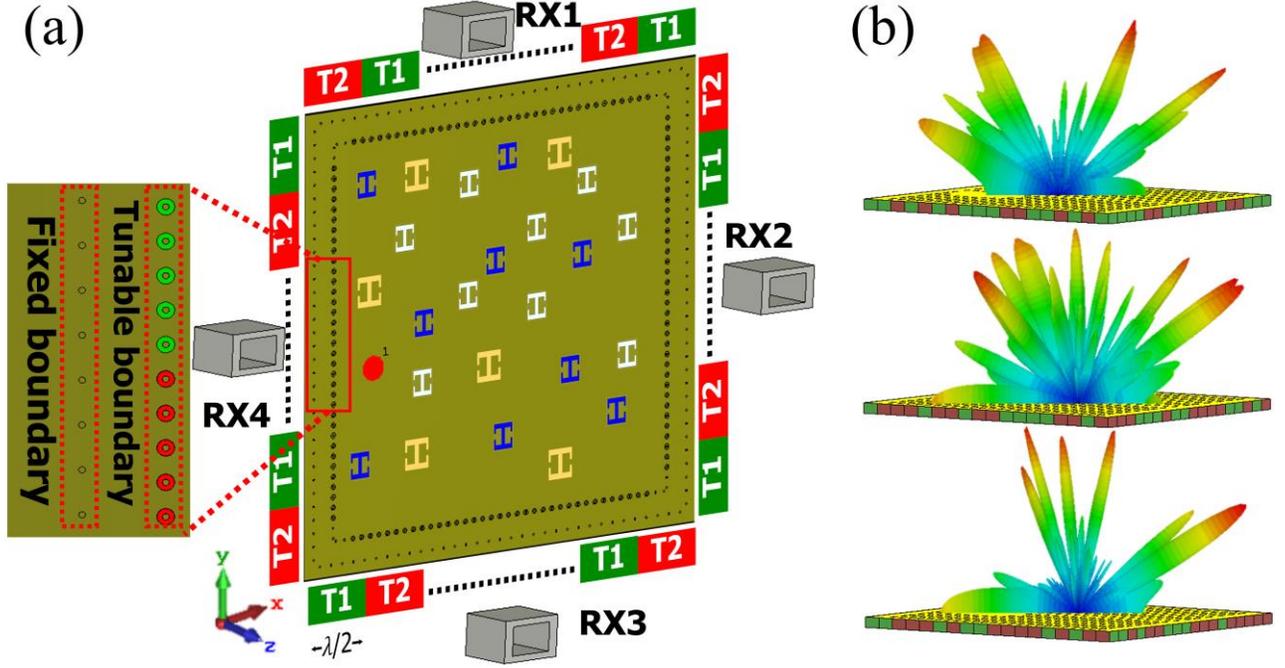

**Fig. 1.** (a) Schematic overview of the imaging system using planar boundary-tunable metasurface antenna, Yellow Blue and White colors indicate designs A, B and C of metamaterial elements. Different boxes around the sides of antenna depict different tunable condition of respective subgroup of inner layer boundary, *Green: on condition (T1), Red: off condition (T2)* (b) Radiation patterns of measurement modes for different boundary configurations (depicted as different combination of green and red subgroups)

TABLE I
DIMENSIONS FOR DIFFERENT DESIGN OF METAMATERIAL ELEMENTS

| Design | Geometric parameters (mm) | | | | |
|---|---|---|---|---|---|
| | $W_{sl}$ | $W_s$ | $W_{gap}$ | $L_s$ | $L_{sl}$ |
| A | 1 | 0.5 | 1 | 1.25 | 5.5 |
| B | 1 | 0.5 | 1 | 1.65 | 6.3 |
| C | 1 | 0.5 | 1 | 2.15 | 7.3 |

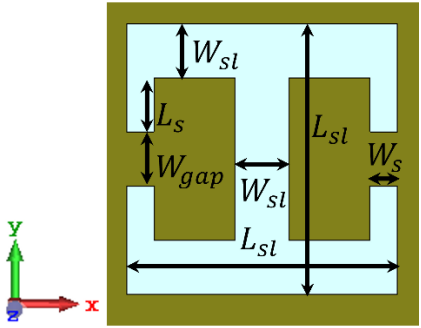

**Fig. 2.** cELC metamaterial elements etched at the top of the antenna. The golden color represents copper and blue represents the opening.

### III. NUMERICAL MODEL OF THE PROPOSED IMAGING SYSTEM

The proposed antenna generates diverse beams which interrogate the scene information to construct an image. The fields generated by the transmitting antenna are reflected from the scene and captured by the receiving antenna. To generalize, in the formulation of the imaging problem it is assumed that the scattered field from the scene can be represented as $E_{scattered} = f(r)E_{incident}$ where f(r) is the spatially varying scene reflectivity. This is the first-order Born approximation which neglects multiple scattering. This approximation leads to fair accuracy [15] for radar imaging and other qualitative security imaging which are the primary intended applications of proposed work. Assuming a specified distance from the antenna, the scene domain is a 2D plane which is pixelated with *N* diffraction limited pixels [47]. To capture the reflected signal, four receiver antennas are modelled at a distance of 6 cm from each side of the transmitting antenna. The whole imaging process is expressed in terms of following equation [47]:

$$g_M = H_{M \times N} f_N \qquad (1)$$

In (1), *g* is the collection of *M* measurements, *H* is referred to as the measurement matrix, and *f* is the unknown reflectivity of the discretized scene of *N* pixels. After applying the Born approximation, the entries of the *H* matrix are represented as $\mathbf{E_{Tx}} \cdot \mathbf{E_{Rx}}$, where $\mathbf{E_{Tx}}$ is the transmitted field of each measurement mode, and $\mathbf{E_{Rx}}$ is the field that would be generated by each receiver if operated in transmission mode.

#### A. Extraction of Polarizabilities

The height of the substrate does not allow any variation of the field along the z direction, since only the fundamental TEM mode can propagate. Under this condition, the scattering within the waveguide cavity becomes a two-dimensional problem. Moreover, the cELC metamaterial elements are subwavelength irises in the top plane of the antenna. The metamaterial elements' scattering response within the cavity can be captured by modelling them as polarizable elements with a small number of multipole moments, typically up to the dipolar terms [48]–[51] as indicated by Bethe's theory [52]. In this section we use the methodology of extracting the *effective polarizability* using the Couple Mode Theory for cylindrical waves adapted from [53]. The *effective polarizability*, ($\alpha^m$) of an element is an environment-dependent quantity, contrary to intrinsic polarizabilities ($\tilde{\alpha}^m$) depending on the geometry of the element [53]. Using CST simulation, the total electric field $E_z^{tot}$ measured at the center of a metamaterial element is extracted by placing it in the center of the parallel-plate waveguide without any other scatterer. Another simulation is performed without the metamaterial element to find the incident electric field, $E_z^{inc}$. The difference between these two quantities gives the scattered electric field $E_z^{sc}$, which can be written as a sum of waveguide modes as follows[53]–[55]:

$$E_z^{sc} = \sum_m \sum_n A_n^{sin} E_{z,sin}^{mn} + A_n^{cos} E_{z,cos}^{mn} \quad (2)$$

In (2) the amplitude coefficients A are calculated by integrating the scattered electric fields on a circle of radius r centered on the metamaterial element with the following equations:

$$A_n^{sin} = \lim_{r \to \infty} \frac{1}{\pi H_n^{(2)}(kr)} \int_0^{2\pi} E_z^{sc}(r,\theta) sin(n\theta) d\theta \quad (3)$$

$$A_n^{cos} = \lim_{r \to \infty} \frac{1}{\pi(1+\delta_{n0}) H_n^{(2)}(kr)} \int_0^{2\pi} E_z^{sc}(r,\theta) cos(n\theta) d\theta \quad (4)$$

The eigenmodes representing the fields are expressed as $E_{z,sin}^{mn} = \frac{\beta_m}{k} H_n^{(2)}(\beta_m r) sin(n\theta)$ and $E_{z,cos}^{mn} = \frac{\beta_m}{k} H_n^{(2)}(\beta_m r) cos(n\theta)$, where $\beta_m = \sqrt{k^2 - (m\pi/h)^2}$ is the propagation constant. To ensure that the evanescent fields do not affect the integration, a radius 55 mm is chosen, satisfying $r \gg \frac{h}{\pi}$, where h is the height of the substrate. Due to the small height of the waveguide only the m = 0 TEM mode propagates. The scattered fields can be expressed as the weighted sum of its eigenmodes through the moments of surface currents, $J_m^n$. Therefore, the scattered fields are written as follows [53],

$$E_z = \frac{m_x Z_0 k^2}{4h} E_{z,sin}^{01} + \frac{m_y Z_0 k^2}{4h} E_{z,cos}^{01} + \frac{p_z k^2}{4ih\epsilon} E_{z,cos}^{00} \quad (5)$$

where, $Z_0$ is the free space wave impedance. By comparing equation (2) and (5) we can find the electric dipole moment and magnetic dipole moments to be, $p_z = A_0^{cos}(4ih\epsilon)/k^2$, $m_y = A_1^{sin}(4h/Z_0 k^2)$, $m_x = A_1^{cos}(4h/Z_0 k^2)$. Using these dipole moments the *effective polarizabilities* within the waveguide are defined as, $\alpha_{zz}^{p,r} = p_z/E_z^{inc}$, $\alpha_{xx}^{m,r} = m_x/H_x^{inc}$ nd $\alpha_{yy}^{m,r} = m_y/H_y^{inc}$. The first subscript of these polarizabilities indicates the direction of resultant dipole moment due the incident field with polarization indicated by the second subscript. These polarizabilities are used in the next section to calculate the induced currents and magnetic dipole moments in the system of equation describing the whole configuration of the proposed antenna. Fig. 3 depicts the comparative overview of the CST exported fields along the circular contour of integration, and analytically calculated scattered fields using equation (5). The relative error due to ignoring the higher order multipole moments of the metamaterial elements is calculated as:

$$\Delta_{sc} = \sqrt{\frac{\int_{\phi=0}^{2\pi} |E_z^{analytical}(\rho=r,\phi) - E_z^{CST}(\rho=r,\phi)|^2}{\int_0^{2\pi} |E_z^{CST}(\rho=r,\phi)|^2}} \quad (6)$$

Here, $E_z^{analytical}$ is the calculated scattered electric field from Eq. (5) whereas $E_z^{CST}$ is the electric field exported from CST simulation. These relative errors are calculated for the three designs of metamaterial elements and are shown in Fig. 4. The relative error is less than 2% over the bandwidth demonstrating accuracy of the dipole model.

*B. Semi-analytical Model of Proposed Antenna*

The antenna is modelled semi-analytically, representing the vias and metamaterial elements by their polarizability, and coupling them via the fundamental waveguide mode. In this model we modelled the co-ordinate system aligned with the center of the antenna allowing the radiation of the antenna in +z direction. The antenna spans from -250 mm to 250 mm in both x and y direction. This model allows us to calculate measurement modes much faster than full wave CST simulation. A current source representing the feed is placed inside the waveguide (the red dot in Fig. 1(a)), generating cylindrical wavefronts inside the antenna. The metamaterial elements are modelled as magnetic dipole moments ($m_x^r$ and $m_y^r$) and electric current ($I_z^r$). Coupling of metamaterial elements through free space radiation is neglected. The vias are modelled by their electric dipole response which is captured by modelling them as electric currents $I_z^v$. We have used the following system of equation (adapted from [48]) for calculating the induced magnetic dipole moments and currents in the metamaterial elements and vias:

$$\begin{bmatrix} Y_{xx}^{r,r} & Y_{xy}^{r,r} & A_{xz}^{r,r} & T_{xz}^{r,v} \\ Y_{yx}^{r,r} & Y_{yy}^{r,r} & A_{yz}^{r,r} & T_{yz}^{r,v} \\ B_{zx}^{r,r} & B_{zy}^{r,r} & Z_{zz}^{r,r} & Z_{zz}^{r,v} \\ G_{zx}^{v,r} & G_{zy}^{v,r} & Z_{zz}^{v,r} & Z_{zz}^{v,v} \end{bmatrix} \begin{bmatrix} m_x^r \\ m_y^r \\ I_z^r \\ I_z^v \end{bmatrix} = \begin{bmatrix} H_x^{r,s} \\ H_y^{r,s} \\ E_z^{r,s} \\ E_z^{v,s} \end{bmatrix} \quad (7)$$

In (7) the superscripts r, v and s represent the radiator, via and the source respectively. The first super/subscript indicates the observation location whereas the second super/subscript indicates the source location. The off-diagonal entries of this matrix are composed of Green's functions, which are presented in Table II. The inverse of extracted magnetic polarizabilities $1/\alpha_{xx}^{m,r}$ and $1/\alpha_{yy}^{m,r}$ are used as the self-interaction term in the diagonal entries $Y_{xx}^{r,r}$ and $Y_{yy}^{r,r}$ respectively whereas $j\omega/(\alpha_{zz}^{p,r} h)$ is used in the diagonal entries of $Z_{zz}^{r,r}$. The factor $j\omega/h$ converts the quantities to be suitable for the current based equation. Similarly, $(j\omega\mu)/h$ is included in the definition of $G_{zx}^{em}$ and $G_{zx}^{em}$ to be suitable for the magnetic dipole moment-based equation here. The self-interaction terms of the vias are calculated from the following analytical expression [47]:

$$[Z_{zz}^{v,v}]_{mn} = \frac{k^2}{4\omega\epsilon} H_1^{(2)}(ka) \quad m=n \qquad (8)$$

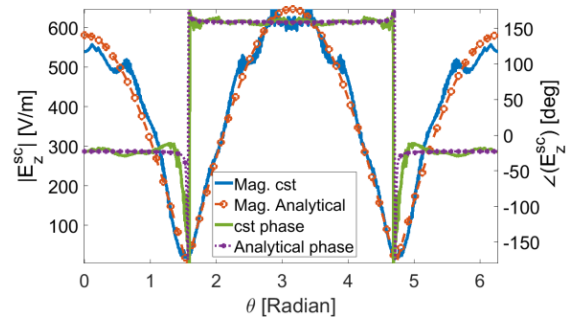

**Fig. 3.** Comparison of scattered field inside the waveguide cavity, analytically calculated $E_z^{sc}$ field for 11.2 GHz along the circular contour of 55 mm with equation (5) in contrast to full wave CST simulation.

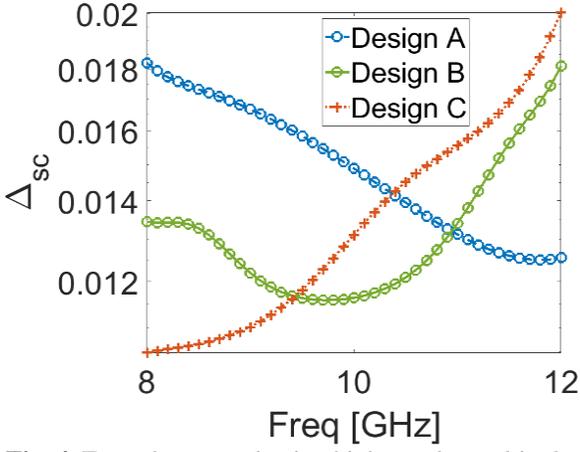

**Fig. 4.** Error due to neglecting higher order multipole moments term for different designs of unit cell over the frequency range.

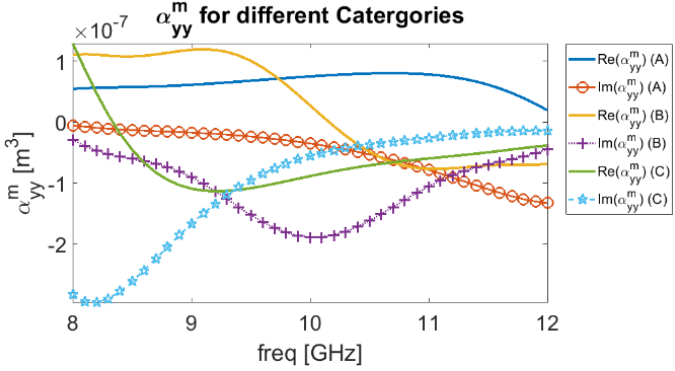

**Fig. 5.** Effective magnetic polarizabilities $\alpha_{yy}^m$ for designs A, B and C, resonating at 8 GHz, 10 GHz, and 12 GHz respectively.

In (8), $H_1^{(2)}(.)$ is the 1st order Hankel function of the second kind. The right-hand side of the equation (7) consists of the incident fields from the source (s) on the metamaterial elements (r) or vias (v). The radiated fields at each of the scene points at location, $\bar{r}$ is calculated by invoking the superposition theorem for fields radiated from n radiators expressed as:

$$E(\bar{r}) = -2i\omega\mu_0 k \sum_n \frac{\exp(ikR)}{4\pi}\left(i - \frac{1}{kR}\right)\frac{\bar{R} \times \bar{m}_n}{R} \qquad (9)$$

where $\bar{R} = \bar{r} - \bar{\rho}_n$, R= $|\bar{r} - \bar{\rho}_n|$, $\bar{\rho}_n$ is the location of $n^{th}$ metamaterial radiator and $k$ is the free space wavenumber. Equation (9) is used for generating diverse measurement modes in next section.

TABLE II
GREEN'S FUNCTIONS INSIDE WAVEGUIDE CAVITY

| For Y, B and G entries | For A, T, and Z entries |
|---|---|
| $G_{xx}^{mm} = G_{yy}^{mm} = -\frac{ikZ_0\omega\epsilon}{8h}\left[H_0^{(2)}(kr) + \cos(2\phi)H_2^{(2)}(kr)\right]$ $G_{xy}^{mm} = G_{yx}^{mm}$ $= -\frac{ikZ_0\omega\epsilon}{8h}H_2^{(2)}(kr)sin(2\phi)$ | $G_{xz}^{me} = -\frac{k}{4i}H_1^{(2)}(kr)\sin(\phi)$ |
| $G_{zx}^{em} = -\frac{i\omega\mu}{h}\frac{k}{4i}H_1^{(2)}(kr)sin(\phi)$ $= -\frac{k\omega\mu}{4h}H_1^{(2)}(kr)sin(\phi)$ | $G_{yz}^{me} = \frac{k}{4i}H_1^{(2)}(kr)\cos(\phi)$ |
| $G_{zy}^{em} = \frac{k\omega\mu}{4h}H_1^{(2)}(kr)\cos(\phi)$ | $G_{zz}^{ee} = -\frac{k^2}{4\omega\epsilon}H_0^{(2)}(kr)$ |

### C. Computation of measurement modes

The imaging setup consists of the antenna described in section II-A as a transmitter and four receivers at coordinate (310, 0,0), (-310, 0,0), (0, 310,0) and (0, -310, 0) (in mm) along the four sides of the transmitting antenna. The receiver antennas are modelled as electric dipoles $p_d$ oriented along the x axis and their radiation pattern is modelled with dyadic Greens's function $\overline{\overline{G_{ee}}}(r_{rx}, r_{scene})$ as follows:

$$E_{rx} = \overline{\overline{G_{ee}}}(r_{rx}, r_{scene})\overline{p}_d \qquad (10)$$

In (10), the Green's function is defined as $\overline{\overline{G_{ee}}}(r_{rx}, r_{scene}) = \frac{\mu_0 k^2 \exp(ikR)}{4\pi\epsilon_0 R}\left[\left(1 + \frac{ikR-1}{k^2R^2}\right)\overline{\overline{I}} + \frac{3-3ikR-k^2R^2}{k^2R^2}\frac{RR}{R^2}\right]$, where ; $\overline{R} = \overline{r}_{scene} - \overline{r}_{rx}$. To create different boundary combinations, different subgroups of the tunable boundary are chosen to be turned *on* or *off*. To mimic this behavior in MATLAB we have replaced the on (off) state by the presence (absence) of the respective subgroups of inner vias. Solving (7) and (9) for all $n_t$ boundary configurations and total $n_f$ given frequency points give arise to a total of $N_t \times N_f \times 4$ measurements.

### D. Scene and resolution

The spatial resolution of the imaging system is derived from k-space arguments and is defined by two parameters: a) bandwidth and b) electrical size of the antenna [41]. The bandwidth of the imaging system generally dictates the range resolution while the antenna size defines the range resolution. Our antenna system has an aperture size of 50 cm (16.67 λ) along xy plane and the receiving antennas are situated at 6 cm (2λ) along each quadrant. A cross-range resolution of 29.03 mm can be found from the following formula [12]:

$$\delta_{cr} = \frac{R\lambda}{D} \qquad (11)$$

center frequency wavelength is used in (11). Moreover, with 21 frequency points in the band of 8-12 GHz corresponds to a maximum unambiguous range ($R_{un}$) of 1.5 m according to the formula $R_{un}=c/\Delta f$, $\Delta f$ the step size for the frequency points (0.2 GHz). Reflections from objects beyond this range will create aliasing in the image. To remain within the unambiguous range, a scene at 0.6 m distance in range direction (+z) spanning -0.3m to 0.3 m in x and y direction is chosen. The area of the imaging has been discretized with a step size of 15 mm (λ/2) in both x and y direction. This gives us 1681 pixels in the scene to be reconstructed. In different imaging methods discussed in section IV, this number will be used for measuring the compression ratio as $\left(1-\frac{\text{Number of measurements}}{\text{number of pixels}}\right) \times 100\ \%)$

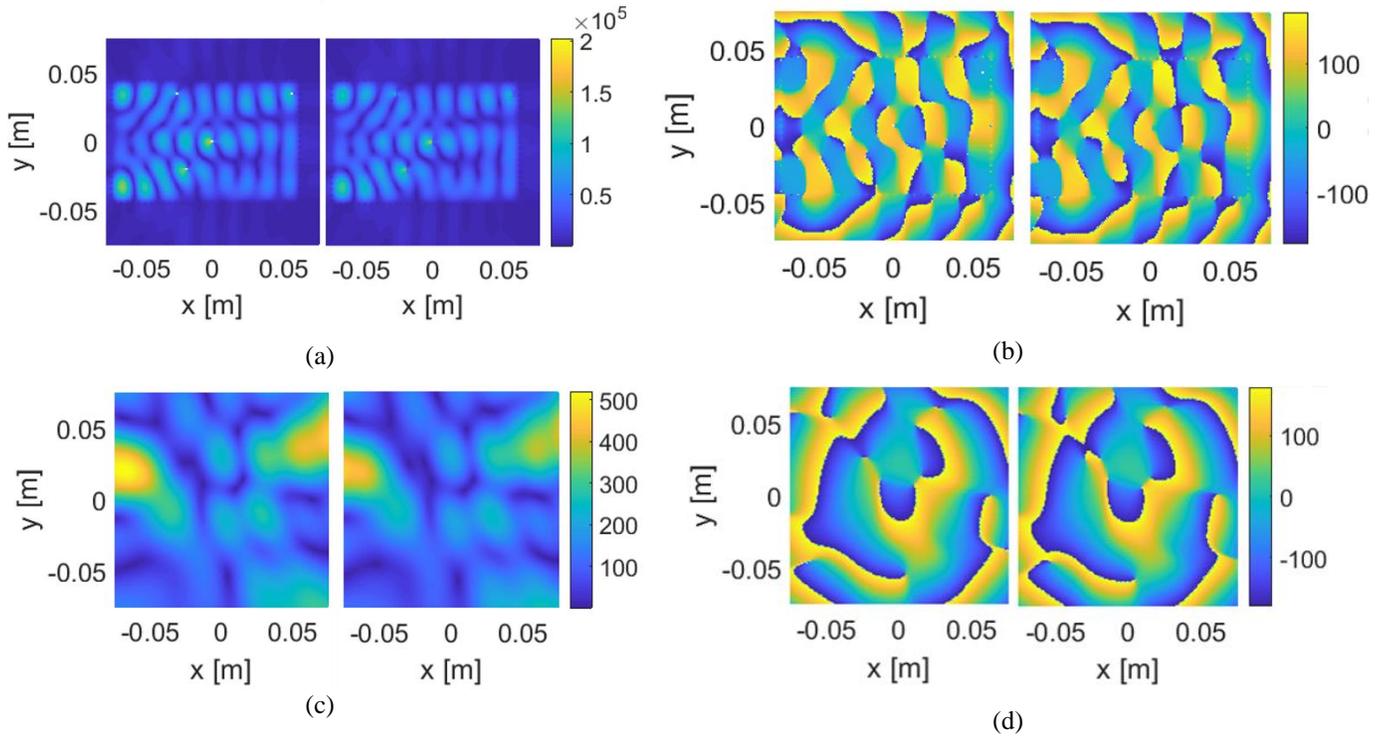

**Fig. 6.** Comparison of numerical and CST generated $E_z$ for 11.2 GHz, (1st column: from semi-analytical model, 2nd column: CST generated field. (a), (b): magnitude and phase for fields inside parallel plate waveguide. (c), (d): magnitude and phase of radiated fields at 80 mm from the antenna.

## IV. RESULTS

Before proceeding with the application of the semi-analytical model to imaging applications, we validate the results of the semi-analytical model against full-wave numerical results produced by CST Microwave Studio. For this purpose, a smaller sized antenna is simulated in CST with a dimension of 75 mm × 75 mm with four identical metamaterial elements of design A as radiators. A similar configuration is also simulated using the semi-analytical model described using Eq. (7) for checking the fields inside the waveguide, while the radiated fields are checked using Eq. (9). The fields inside the waveguide as well as radiated fields calculated with semi-analytical model are compared to CST generated fields Fig 6. For fields inside and radiated from the cavity, the main characteristics are significantly captured by the analytical model. Using a similar relative root mean squared error calculation metric as described in Eq. (6) the errors in the in the case of Fig. 6 (a) and Fig. 6 (b) are $1.72 \times 10^{-3}$ and $1.26 \times 10^{-5}$ respectively, and $3.72 \times 10^{-8}$ and $1.29 \times 10^{-9}$ in the case of Fig. 6(c) and 6(d) respectively. The accuracy of the analytical model is sufficient for quantifying the efficacy of the proposed imaging method.

### A. Comparison of Frequency Diverse Approach vs Boundary Tunable Approach

After validating the semi-analytical model, we apply it for calculating the fields at the scene defined in section III-D with the proposed approach. The radiated fields of different measurement modes are presented in Fig. 7. The imaging process involves solving equation (1) for the scene reflectivity *f*. The ability to resolve the image depends on the rank of the measurement matrix H. A completely determined measurement matrix will have a rank equal to the scene dimension [56] and therefore a simple inverse can solve the matrix problem. However, in most cases the measurement matrix has a smaller number of rows (*measurements*) than columns (or number elements in f), or the measurements have some degree of correlation among them. This leads to a poorly conditioned matrix problem and thus regularization techniques are applied to obtain an estimate of f with the following minimization problem [36], [56]:

$$f_{est} = \underset{f}{\text{argmin}} ||g - Hf||^2 + \lambda\, R(f) \qquad (12)$$

Where λ indicates the regularization parameter which determines relative importance between the first and second part of the objective function and R(f) indicates a regularizer. In computational imaging the $l_1$-norm is often used as a regularizer which favors sparse estimates of the scene reflectivity function f. In many cases the scene is naturally sparse in the pixel basis, or it can be expressed in another basis in which it is sparse (such as a wavelet or curvelet basis) [25]. In general, the goal is to achieve the Restricted Isometry Property (RIP) of the H matrix for efficiently retrieving the scene information using computational techniques such as compressive sensing. RIP is normally achieved by appropriate design of the measurements so that they have high diversity.

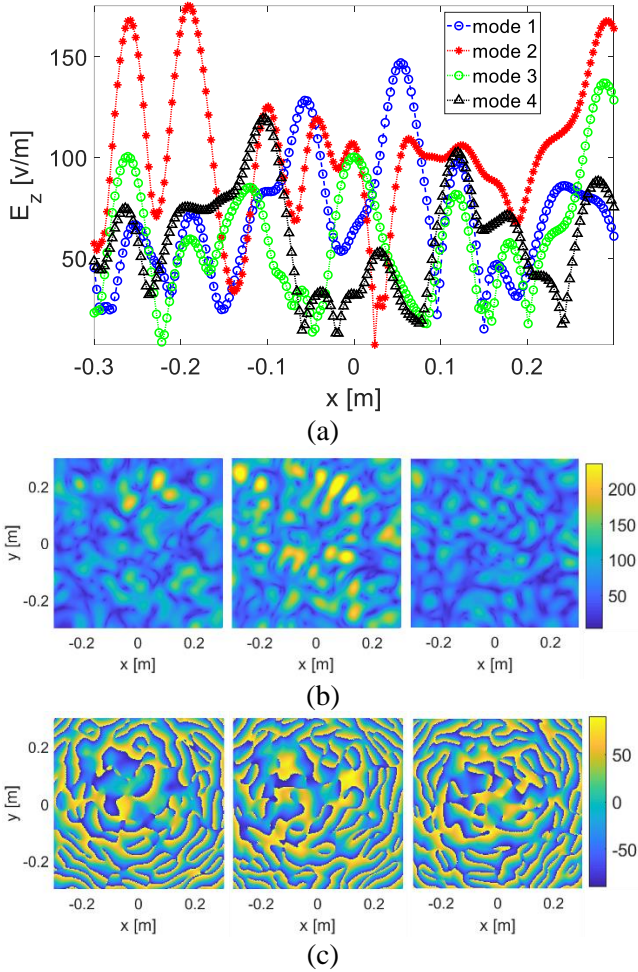

**Fig. 7.** Electric fields for different measurement modes using tunable boundary: (a) $|E_z|$ along y=3 mm line at z=0.6m for four different measurement modes using tunable boundary. (b), (c) amplitude and phase of total field E for three different measurement modes.

A singular value decomposition quantifies the rank and degree of diversity of the measurements captured in H. In the absence of noise, the number of non-zero singular values indicates the rank of the measurement matrix H. In practice, the number of singular values above a specific noise level indicates the *effective rank*. Therefore, robustness to noise and measurement diversity are indicated by a flatter singular value spectrum.

The singular values of matrix H for frequency diverse approach and proposed tunable boundary approach are compared in Fig. 8. In this study the imaging system presented in Fig. 1 is used, with one transmitting antenna and four receiving antennas. The H matrix is numerically calculated with the dot products between the electric fields from transmitter and receiver and then the singular value decomposition is performed. For the frequency diverse approach, different numbers of frequencies (indicated as $n_f$) are used, while for the tunable case different combinations of number of frequencies ($n_f$) and number of tunable boundaries ($n_t$) are used for populating the measurement matrix. In all cases the whole X-band (8-12 GHz) is used for simulating the measurements. In general, the tunable approach gives a higher degree of diversity, as indicated by the flatter singular value spectrum. Increasing the number of frequency points ($n_f$) from 51 to 201 does not increase the diversity much, as indicated by similar slops of the curves. On the other hand, 10 boundary combinations ($n_t$) with only 21 frequency points ($n_f$) outperforms the frequency diverse approach with a similar number of measurements. Moreover, increasing the number of boundary combinations to $n_t$ =20 leads to a slight increase of diversity, as indicated by the red curve. The singular value spectrum analysis of this section is referred to in Section IV C-3, where we discuss the efficacy of our proposed approach at different noise levels.

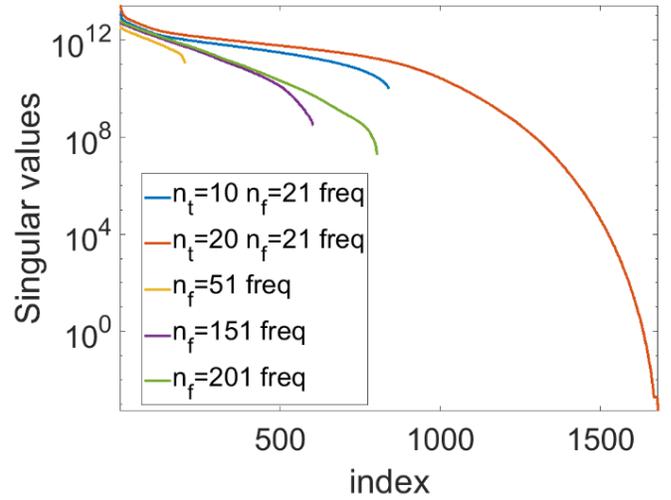

**Fig. 8.** Comparison of Singular Values of H matrix using proposed tunable boundary condition and frequency diverse approach.

*B. Effect of Bandwidth on singular value spectrum*

To investigate the effect of bandwidth on the singular value spectrum, we have simulated the H matrix for smaller bandwidth (8-10 GHz) while maintaining the same number of frequency points (21). As depicted in Fig. 9 (a), smaller bandwidth (shown by the red curve) results in more correlated measurement modes, indicated by faster decaying curve compared to the higher bandwidth case (the blue curve). This indicates that boundary tunability alone cannot outperform the frequency diversity approach. To validate this claim we have simulated the extreme case by simulating a single frequency with 20 boundary combinations and compared it with a frequency diverse approach with 21 frequencies in 8-12 GHz bandwidth. The results are shown in Figure 9 (b). Even though the tunable boundary has higher dominant singular values, it soon decays to lower values. This confirms that combining boundary and frequency tuning maximizes the diversity of the measurement modes.

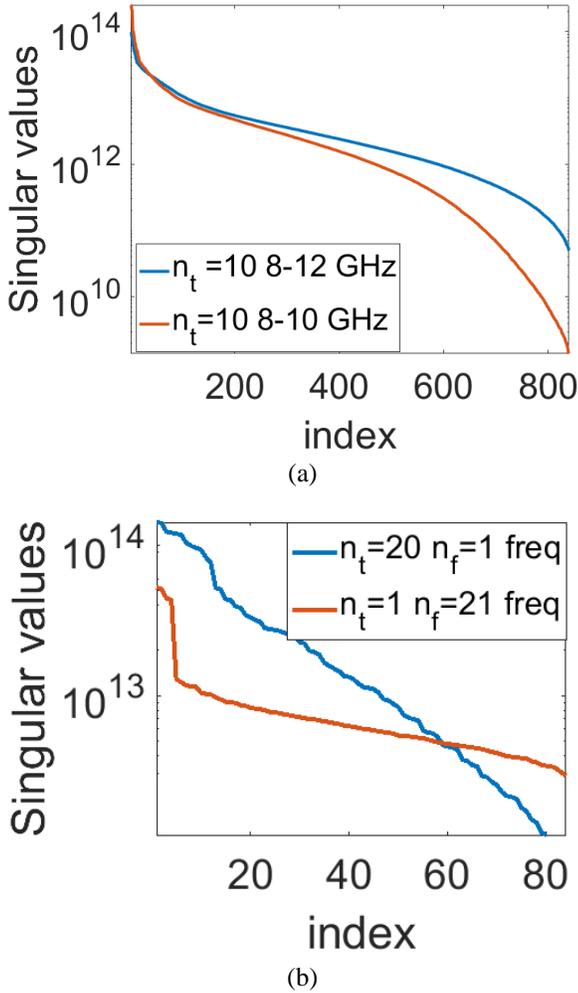

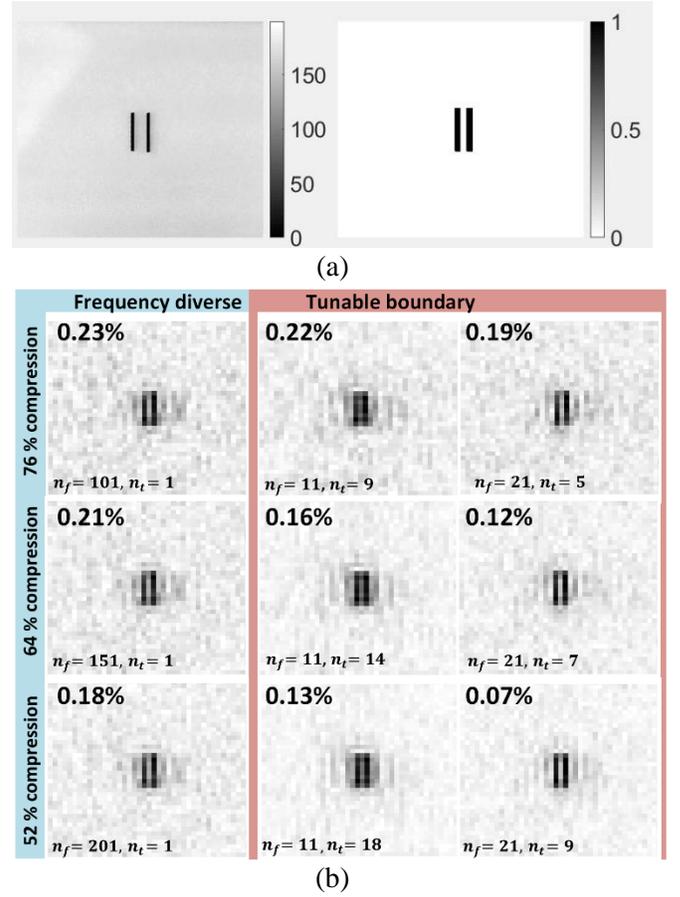

**Fig. 9.** (a) Effect of Bandwidth on singular value spectrum for the tunable boundary approach with 21 frequency points. (b) Singular value spectrum for single frequency and single boundary approach.

*C. Imaging performance*

   *1. Definition of the scene*

We use our semi-analytical model to compare the performance of our boundary tunable approach with the frequency diverse approach. Two different scenes are used: i) Parallel stripes scene and ii) Multi-patch scene. *Parallel stripes scene* consists of two small coaxial cable of 5 cm × 1 cm placed 1 cm apart. *Multi-patch scene* consists of two L-shaped objects (with longer arm of 5 cm × 2.5 cm and 4 cm × 2 cm, and with side arm of 2.5 cm × 2.5 cm and 2 cm × 2 cm respectively), two inclined strips (5 cm × 2 cm) and 5 cm ×1 cm) and a square (4 cm × 4 cm) in the center of the scene. In this setup we used pictures taken by a mobile phone camera placed from at 0.6 m from a scene spanning -0.3 m ≤x≤0.3 m and -0.3 m ≤y≤0.3 m and then down sampled with a resolution of 15 mm ($\lambda/2$). The obtained images are converted into black and white using thresholding to obtain a sparse representation in the pixel basis. The pictures of these scenes taken by the camera are shown in Fig. 10 (a) and Fig. 11 (a) in the left column respectively, while the black and white images used to model the scene are shown in the right column.

**Fig. 10.** Scene and recovered image of parallel stripes scene. (a) The image captured by phone (left) and down sampled black and white input image (right). (b): different images retrieved by different levels of compression with the frequency diverse approach (*left column*), and the tunable boundary approach with $n_f$ =11 frequencies (middle column) and $n_f$ =21 frequencies (right column) with varied boundary configurations $n_t$. Mean squared error is also shown for each figure. The scaling of each image is same as in (a).

   *2. Role of tunable boundaries and frequencies:*

In this study we compare the frequency diverse approach and the tunable boundary approach using *parallel stripes scene*. It is worth noting that, in the tunable boundary approach, to generate a similar number of measurements as in the frequency diverse approach, we have two design knobs namely the number of frequencies ($n_f$) and the number of tunable boundaries ($n_t$). The total number of measurements in the tunable boundary approach is $n_f \times n_t \times 4$, where the factor 4 is due to the four receivers. Therefore, a similar number of measurements can be generated using a higher number of frequencies and a lower number of tunable boundaries and vice-versa. In this section the main objective is to address the following questions:

I. Does the tunable boundary approach lead to higher imaging performance than frequency diverse approach with respect to a common ground truth?

II. Which combinations of tunable boundaries and frequencies leads to the highest imaging performance?

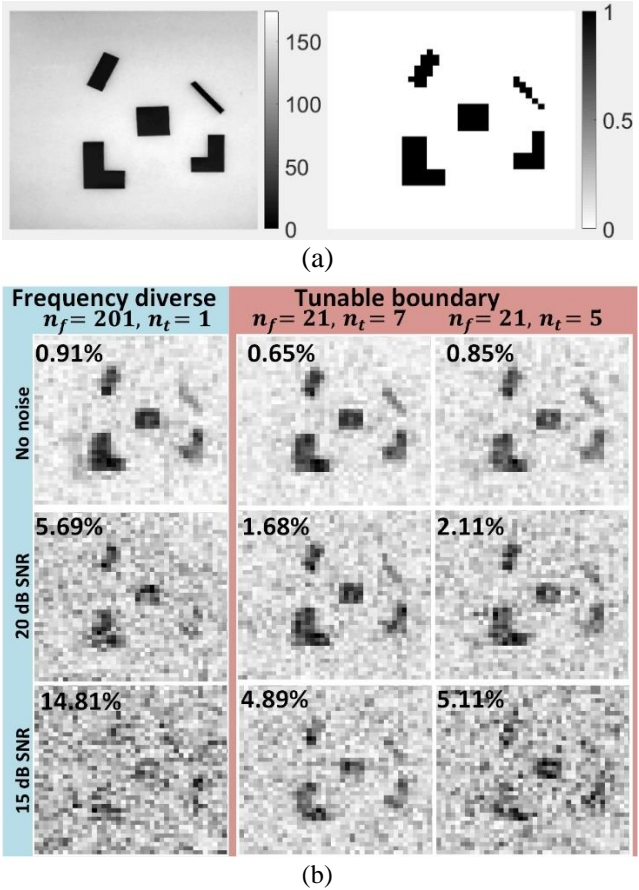

**Fig. 11.** Comparison of frequency diverse approach and tunable boundary approach with respect to robustness against noise: (a) The image captured by phone (left) and down sampled black and white input image (right) of the multi-patch scene. (b) images retrieved with different SNR levels. Mean squared errors (in percentage) are shown in the insets. The scaling of each image is same as in (a).

To answer these questions, we perform synthetic imaging of the *parallel stripes scene* using the frequency diverse approach with $n_f$=101, 151 and 201 frequencies resulting 404, 604 and 804 total measurements respectively. Compared to 1681 number of pixels these measurements refer to an imaging system with 76%, 64% and 52% compression respectively. On the other hand, in the boundary tuning approach, the imaging is performed using 11 frequencies $(n_f = 11)$ and 21 frequencies $(n_f = 21)$ with varied number of tunable boundaries $(n_t)$ while maintaining a similar number of total measurements as in the case of frequency tunable approach. The imaging is performed by solving the minimization problem of Eq. (12) with the Two-step Iterative Shrinkage and Thresholding (TwIST) [46] algorithm using the $l_1$ − norm of the scene reflectivity as the regularizer (R(f)). The retrieved images are compared in Fig. 10 (b). The retrieved image using a fully determined H matrix with $n_f = 21$ and $n_t$=20 boundaries is used as the ground truth for obtaining the mean squared error for each image. These errors for different images are shown in the inset of the figures (at top left corner). In general, the tunable boundary approach performs better in imaging as depicted by the lower mean squared errors.

Comparing images obtained by tunable boundaries (second and third column of Fig. 10 (b)), we can observe that a higher number of frequencies ($n_f = 21$) leads to better imaging than with a lower number of frequencies ($n_f = 11$). For this reason, we have utilized $n_f = 21$ in the following study of robustness of the proposed approach against noise.

*3. Performance analysis in the presence of noise*

In this study *multi-patch scene* is imaged with different levels of added white Gaussian noise. It is worth noting that while *parallel stripes scene* served as a tool to contrast the frequency diverse approach and proposed approach in noise free scenario, this *multi-patch scene* will serve as a tool for comparing their ability of complex scene imaging in noisy scenario due to the different shapes and orientations of the objects in the scene. In this case we use 52% compression ($n_f = 201$, total measurements of 804) for the frequency diverse approach and 65% and 75% compression ($n_f = 21\ with\ n_t = 7$, and $n_f = 21\ with\ n_t = 5$, or total of 588 and 420 measurements respectively) for the tunable boundary approach. We have indicated the mean squared error compared to the ground truth image formed using $n_f = 21$ and $n_t$=20 without any noise. The images are shown in Fig. 11 (b), for cases with no noise, 20 dB SNR and 15 dB SNR with the mean squared errors indicated in the insets. The quality of the images obtained with the tunable boundary approach with higher compression (lesser number of measurements) is better than their counterparts obtained with the frequency diverse approach having lesser compression (higher number of measurements), as depicted by lower root mean square errors. For 15 dB SNR, the frequency diverse approach exhibits a mean squared error of 14.81% with 52% compression whereas the tunable boundary approach exhibits 4.9% and 5.1% error with compression of 65% and 75% respectively.

Here we would like to revisit the singular value spectrum shown in Fig. 8. As discussed in section IV-A, due to the flatter (slowly decaying) spectrum exhibited by tunable boundary approach, it is more robust against noise. The lower mean squared errors for different noise levels are consistent with the flatter singular value spectrum exhibited by tunable boundary approach in contrast with the frequency diverse approach.

## V. CONCLUSION

A metasurface antenna is proposed, based on a boundary tunable parallel plate waveguide cavity. A semi-analytical model is used for efficiently simulating the imaging system in MATLAB. Images of different scenes retrieved with the TwIST algorithm proved the efficacy of the proposed antenna as well as its robustness against noise down to 15 dB SNR. Using other regularizers such as total variation [15] may lead to further improvement in imaging performance. Moreover, generating intelligent measurement modes for task specific imaging has been shown to be promising, [30], [57]–[60], and could be performed with our proposed structure. With the inclusion of higher order multipole moments [54], [55], the accuracy of the semi-analytical model could be improved. This would enable the model to be used for quantitative imaging requirements, reconfigurable antenna design and the recently developed concept of smart radio environments [61], [62] for wireless communication in 6G applications.

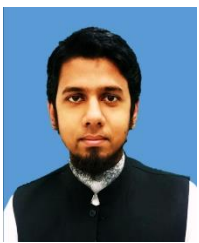

**Toufiq Md Hossain** (Graduate Student Member, IEEE), is a PhD student in School of Engineering & Information Technology in University of New South Wales, Canberra. Prior to this he was research assistant at Advanced Communication Engineering Centre (ACE) in Universiti Malaysia Perlis and completed his M.Sc. in Communication Engineering from the same university. He has received his B.Sc. in Electrical and Electronic Engineering (EEE) from Khulna University of Engineering & Technology (KUET), Khulna, in 2014. He is recipient of TFS scholarship from UNSW in 2019 for his PhD. He has authored or co-authored 12 refereed journals and 13 international conferences. He is a graduate student member of IEEE antennas and propagation society and IEEE Microwave Theory and Techniques Society. His research interests include Metamaterials, Metasurfaces based computational microwave imaging, flexible linear to circular polarizers for CubeSat applications.

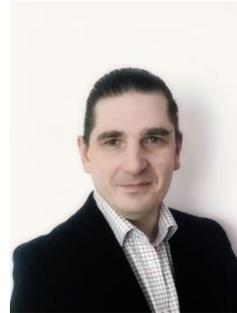

Prof Miroshnichenko obtained his Ph.D. in 2003 from the Max-Planck Institute for Physics of Complex Systems in Dresden, Germany. In 2004 he moved to Australia to join the Nonlinear Physics Centre at ANU. During that time, he made fundamentally important contributions to the field of photonic crystals and brought the concept of the Fano resonances to nanophotonics. In 2007 he was awarded by Australian Postdoctoral and in 2011 by Future Fellowships from the Australian Research Council. In 2017 he moved to the University of New South Wales Canberra and got a very prestigious UNSW Scientia Fellowship. During his career, Prof Miroshnichenko published more than 250 journal publications and has authored a book, in addition to writing 7 book chapters in other publications. He was recognized as a 'Highly Cited Researcher' for three consecutive years (2019-2021) by the Web of Science Group, which is awarded to fewer than 0.1% of the world's most influential researchers. Recently, he was also recognized as a top-40 researcher in Australia (and top-5 in Physics) by The Australian Lifetime Achievers Leaderboard (2020, 2021). The topics of his research are nonlinear nanophotonics, nonlinear optics, and resonant interaction of light with nanoclusters, including optical nanoantennas and metamaterials.

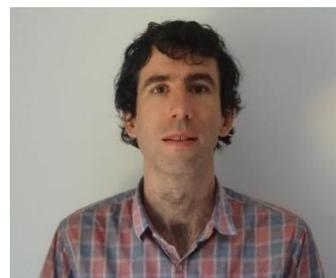

**David A. Powell** (Senior Member, IEEE) is a Senior Lecturer with the School of Engineering and Information Technology, University of New South Wales, Canberra, Australia. His is a Senior Member, of the IEEE, and has been a member since 2000. Since 2018 he has served as chair of the Australian Capital Territory IEEE AP/MTT joint chapter, and he has been a Senior Member of URSI since 2020. Between 2006 and 2017, he was a Researcher with the Nonlinear Physics Centre, The Australian National University, Canberra. He received the Ph.D. degree in Electronic and Communications Engineering from RMIT University, Melbourne, Australia in 2006 and the Bachelor of Computer Science and Engineering degree from Monash University, Melbourne, Australia, in 2001.

Dr Powell has co-authored 4 book chapters, and 87 articles in refereed international journals, and has delivered 12 invited and 1 keynote talk at international conferences. He has successfully graduated 8 PhD students and 1 Masters by Research student, and is currently supervising 3 PhD students. He is an inventor or contributor to 3 patents, and developed an open-source code OpenModes for finding the modes of metamaterial structures. He has been a Chief Investigator on 6 successfully funded grants from the Australian Research Council, as well as 1 grant from the New South Wales Defence Innovation Network. He has acted as a reviewer for over 30 different international journals in physics and engineering.

He has conducted research in electromagnetic metamaterials for microwave, millimeter-wave, terahertz, and near-infrared wavelength ranges, in addition acoustic metamaterials. His interests include metasurfaces, nonlinear and tunable metamaterials, bianisotropy, chirality, near-field interaction effects, acoustic radiation forces and achromatic structures.